# Observation of Shubnikov-de Haas Oscillations in Large-Scale Weyl Semimetal WTe$_2$ Films

Yequan Chen, Yongda Chen, Jiai Ning, Liming Chen, Wenzhuo Zhuang, Liang He, Rong Zhang, Yongbing Xu[*], Xuefeng Wang[*]

Jiangsu Provincial Key Laboratory of Advanced Photonic and Electronic Materials, School of Electronic Science and Engineering, and Collaborative Innovation Center of Advanced Microstructures, Nanjing University, Nanjing 210093, China

[*]Corresponding Authors. Email: xfwang@nju.edu.cn (X.W.); ybxu@nju.edu.cn (Y.X.)

**Abstract:** Topological Weyl semimetal WTe$_2$ with large-scale film form has a promising prospect for new-generation spintronic devices. However, it remains a hard task to suppress the defect states in large-scale WTe$_2$ films due to the chemical nature. Here, we significantly improve the crystalline quality and remove the Te vacancies in WTe$_2$ films by post annealing. We observe the distinct Shubnikov-de Haas quantum oscillations in WTe$_2$ films. The nontrivial Berry phase can be revealed by Landau fan diagram analysis. The Hall mobility of WTe$_2$ films can reach 1245 cm$^2$V$^{-1}$s$^{-1}$ and 1423 cm$^2$V$^{-1}$s$^{-1}$ for holes and electrons with the carrier density of 5 × 10$^{19}$ cm$^{-3}$ and 2 × 10$^{19}$ cm$^{-3}$, respectively. Our work provides a feasible route to obtain high-quality Weyl semimetal films for the future topological quantum device applications.

Shubnikov-de Haas (SdH) quantum oscillations in conductivity has become one of the most effective tools to explore novel physics in metals, semimetals and narrow-gap semiconductors since it was firstly discovered in bismuth in 1930.[1] It poses a way to describe the transport properties of related carriers close to Fermi surface.[2,3] Recently, the burgeoning topological quantum materials, such as topological insulators, Dirac semimetals and Weyl semimetals, have established a new state of condensed matter.[4-6] Unique electronic structures and strong spin-orbit coupling around Fermi surface bring them exotic physical properties, including unusually large magnetoresistance (MR)[7] and new nontrivial topological electronic state.[8] By analysis of SdH oscillations in these topological quantum materials, such as $Bi_2Se_3$[9] and $Cd_3As_2$,[10] the transport properties of relevant quasiparticles and their closed orbits at the Fermi surfaces can be thoroughly clarified.

$WTe_2$, as a typical Weyl semimetal material, has type-II Weyl points resulted from the lack of inversion symmetry. It possesses a layered structure with an additional structural distortion along the crystallographic axis *a* of tungsten chain.[11-13] Experimentally, it shows extremely unsaturated large MR of ~13,000,000% in magnetic fields up to 60 T at 0.53 K.[11] Moreover, it has been demonstrated to own affluent Weyl features in three dimensions[14-17] and alluring topological quantum state in two dimensions.[18-22] Therefore, to elucidate the origin of these attractive physical properties, indispensable dissection based on the observed SdH oscillations is frequently implemented. Integration into the potential applications of spintronic devices and quantum computer, fabrication of smooth and continuous $WTe_2$ films is the first and very crucial step. The related fabrication techniques are chemical vapor deposition,[23-26] molecular beam epitaxy,[19, 27, 28] and pulsed laser deposition (PLD).[29-31] Despite great efforts have been devoted to achieving large-scale $WTe_2$ films, obtaining high-quality samples is of great difficulty due to the very low vapor pressure of cation tungsten. This limit prevents the observation of the typical SdH oscillations and/or other quantum effects in large-scale $WTe_2$ films. Thus, it still remains an urgent task to map the Fermi surface and its topology in continuously large-scale WTe$_2$ films.

In this Letter, we fabricate the high-quality, large-scale, single-crystalline Weyl semimetal WTe$_2$ films on mica substrates by PLD and the post-annealing method. The SdH oscillations are observed in these WTe$_2$ films. According to the fast Fourier transform (FFT) and analysis of SdH quantum oscillations, the effective mass, the nontrivial Berry phase, the Dingle temperature, and the quantum mobility are determined. In addition, through two-carriers model fitting, the Hall mobility up to 1423 cm$^2$V$^{-1}$s$^{-1}$ is obtained, while the lower carrier density is achieved as compared to our previous work.[30]

The WTe$_2$ target was prepared by the melting method.[30] The basic pressure of the PLD vacuum chamber was ~2.9×10$^{-5}$ Pa and the distance between the substrate and the WTe$_2$ target was ~5 cm, as shown in Fig. 1(a). The amorphous WTe$_2$ films were deposited onto the substrate at 300ºC at the speed of 1 nm/min for 100 min using a 248 nm KrF excimer laser beam (an average fluence of 1 J/cm$^2$ and a repetition rate of 2 Hz). Subsequently, an annealing process was implemented to crystallize the amorphous WTe$_2$ films. The films were sealed in a 20 mL evacuated quartz tube with 3.3 mg Te powder, and then were annealed at 700ºC for 48, 72 and 96 hours, respectively. Note that other annealing temperatures turned out the worse samples (see Fig. S1 in the Supplementary Material). The crystalline structure of these WTe$_2$ films was measured by a micro-Raman spectrometer (NT-MDT nanofinder-30) with a 514.5 nm Ar$^+$ laser and x-ray diffraction (XRD) using a Cu K$\alpha$ line (Rigaku Ultima III). Selected area electron diffraction (SAED) attached in a transmission electron microscopy was used to confirm the single-crystalline nature. The thickness and surface morphology were examined by the atomic force microscope (AFM) system (Asylum Cypher) with the results shown in Fig. S2 in the Supplementary Material. Furthermore, the films were cut into the size of 2.5×5×0.2 mm$^3$ for the transport measurements, which was carried out in the standard Hall and resistivity configuration at low temperatures with a field up to 14 T (Oxford-14 T). Standard lock-in amplifiers (Stanford Research System SR830) with a low frequency (<20 Hz) excitation current of 1 μA (Keithley 6221) were used.

Figure 1(b) shows the Raman spectra of the WTe$_2$ films, which reveals seven main phonon peaks at about 79, 88, 110, 115, 132, 162, and 120 cm$^{-1}$, corresponding to $A_1^1$, $A_2^3$, $A_2^4$, $A_1^9$, $A_1^8$, $A_1^5$ and $A_1^2$ vibrational modes,[32] respectively. The XRD patterns exhibit diffraction peaks corresponding to (00*l*) family of planes [Fig. 1(c)], indicating the cleaved surface perpendicular to the *c* axis of the WTe$_2$ films.[33] Notably, the XRD pattern of the WTe$_2$ films annealed for 48 hours has the widest full width at half maximum. In addition, the indiscernible diffraction peaks of (004) and (0010) for the 48-hour-annealed WTe$_2$ film indicate that the relatively fast annealing time (48 hours) suppresses the crystallization process. Furthermore, the SAED pattern further corroborates the single-crystalline nature of the WTe$_2$ film annealed for 72 hours [Fig. 1(d)].

The temperature-dependent longitudinal resistivity ($\rho_{xx}$) of WTe$_2$ films is displayed in Fig. 2(a), which is normalized by their values at 300 K. It is well-known that single-crystalline bulk WTe$_2$ has the well-defined metallic behavior.[11] Here, all the films show the metallic behavior beyond 15 K. Below 15 K, the low-temperature resistivity minima are visible, which are mainly attributed to the weak localization in quantum interference effects.[30] The residual resistivity ratios (RRRs, defined as RRR = $\frac{\rho_{300K}}{\rho_{1.7K}}$) of WTe$_2$ film annealed for 48, 72 and 96 hours are 1.75, 7.27 and 1.44, respectively. The RRR of WTe$_2$ film annealed for 96 hours is unexpectedly smaller than others, suggesting that the long annealing time may cause the existence of the much more content of Te vacancies in the WTe$_2$ films.[23, 34] The MR curves (MR = $\frac{\rho(B)-\rho_0}{\rho_0} \times 100\%$) are shown in Fig. 2(b), where the external field is along the *c* axis of the films. The WTe$_2$ film annealed for 72 (96) hours has the maximum (minimum) MR value of ~ 321% (11%) at 14 T. This agrees well with our previous work that the Te deficiency in WTe2 samples dramatically declines the MR value.[30] Moreover, according to the Lorentz law MR ≈ $(\mu_{avg} \cdot B)^2$, the average mobility of the WTe2 films annealed for 48, 72 and 96 hours is calculated to be 621, 1347 and 248 cm$^2$V$^{-1}$s$^{-1}$, respectively. Therefore, the 48- and 96-hour-annealing duration suppresses the values of RRR and MR due to the incomplete crystallization as well as the much more Te vacancies.

To clarify the transport properties of the WTe2 film annealed for 72 hours, the dependence of the MR up to 14 T on temperature (from 1.7 to 100 K) is shown in Fig. 3(a). From 1.7 to 10 K, the MR value increases a little bit due to the faded weak localization component with increasing temperature.[30] Above 10 K, the MR value ordinarily decreases with increasing temperature due to the increased inelastic scattering,[35] i.e., electron-phonon scattering. Remarkably, the clear SdH oscillations periodic in 1/$B$ at temperatures between 1.7 and 10 K are extracted in Fig. 3(b) after a smooth subtraction of background. The FFT spectra of the SdH oscillations at temperatures between 1.7 and 10 K are shown in Fig. 3(c). It reveals a single oscillation frequency $F$ =175.4 T of the SdH oscillations. According to the Onsager relation $F = (\hbar/2\pi e)A_F$, where $\hbar$ is reduced Planck's constant and $A_F$ is the cross-sectional area of this Fermi pocket normal to the field, $A_F$ is determined to be ~0.017 Å$^{-2}$. Furthermore, the Fermi wave vector $k_F$ = 0.074 Å$^{-1}$ can be acquired by a circular cross-sectional approximation. Landau fan diagram is shown in the inset of Fig. 3(c), where the minimum of $-d^2G_{xx}/d^2B$ ($G_{xx}$ is defined as $\frac{R_{xx}}{R_{xx}^2+R_{xy}^2}$) is chosen as the integral number of Landau level. Linear fitting line gives a nonzero intercept of ~0.452, corresponding to a nontrivial Berry phase of π. In addition, the slope of this fitting line can also be described by the semiclassical Onsager equation: $k_{n\sim 1/B}$ = $A_F \cdot \hbar/2\pi e$,[36] which results in the range of Landau fan diagram from 13.5 to 17.

Generally, the SdH quantum oscillations can be described by the Lifshitz-Kosevich (LK) formula. The temperature dependence of the FFT amplitude can be defined as:

$$A_{\text{FFT}} \sim \Delta R \propto [(\lambda \cdot T \cdot m^*/B)/\sinh(\lambda \cdot T \cdot m^*/B)] \qquad (1)$$

where the $A_{\text{FFT}}$ is the FFT amplitude, $\lambda = 2\pi^2 k_B/\hbar e$ ($k_B$ is the Boltzmann constant), $T$ is the temperature, $m^*$ is the effective mass and the $1/B$ is the average inverse field of the Fourier window.[33] In Fig. 3(d), temperature dependence of the FFT amplitudes is normalized by their 1.7 K values. The black fitting line provides the effective mass of ~0.3605 $m_e$, comparable to those of high-quality exfoliated samples.[33] The Fermi velocity $v_F = \hbar k_F/m^*$ and the Fermi energy $E_F = (\hbar k_F)^2/m^*$ are ~2.35 × 10$^5$ m/s and ~113 meV, respectively. The Dingle plot of the $\text{Log}(\Delta R \cdot B \cdot \sinh(\lambda T m^*/B)/R)$ versus the inverse field $1/B$ is shown in Fig. 3(e). According to the slope $k\ln10 = -2\pi^2 k_B T_D m^*/\hbar e$ of the linear fitting line, we can extract the Dingle temperature $T_D$ of ~4.74 K. Thereafter, the quantum scattering time $\tau = \hbar/2\pi k_B T_D$ is calculated to be 2.57 × 10$^{-13}$ s, so that the quantum mobility $\mu_q = e\cdot\tau/m^*$ is ~ 1249 cm$^2$V$^{-1}$s$^{-1}$. A mean free path of $l_q$ = 60.4 nm is estimated by the relation $l_q = v_F\cdot\tau$, which is comparable to those of other topological semimetals.[37-39]

In WTe$_2$, the Fermi level is located at both conduction and valence bands simultaneously, which makes two kinds of carriers (electrons and holes) existing at the Fermi surface.[11] The nonlinear trend of Hall resistivity shown in Fig. 4 further proves the existence of electrons and holes in our WTe$_2$ films. To gain the Hall mobility and density of electrons and holes, the two-carriers model is used to fit the magnetic-field dependent magneto-resistivity $\rho_{xx}$ and Hall resistivity $\rho_{xy}$ at 1.7 K simultaneously:[40]

$$\rho_{xx} = \frac{1}{e} \frac{(p\mu_h + n\mu_e) + (p\mu_h\mu_e^2 + n\mu_e\mu_h^2)B^2}{(p\mu_h + n\mu_e)^2 + (p-n)^2 \mu_h^2 \mu_e^2 B^2} \qquad (2)$$

$$\rho_{xy} = \frac{B}{e} \frac{(p\mu_h^2 - n\mu_e^2) + (p-n)\mu_h^2\mu_e^2 B^2}{(p\mu_h + n\mu_e)^2 + (p-n)^2\mu_h^2\mu_e^2 B^2} \qquad (3)$$

where $p$ and $n$ represent the density of holes and electrons, respectively; $\mu_h$ and $\mu_e$ are the Hall mobility of holes and electrons, respectively. Here, we obtain $p = 5 \times 10^{19}$ cm$^{-3}$ and $n = 2 \times 10^{19}$ cm$^{-3}$. Actually, the density of holes and electrons should be perfectly compensated in WTe$_2$. However, considering the existence of the Te vacancies, it is expected that there is a difference between the density of holes and electrons. The $\mu_h$ and $\mu_e$ are calculated to be ~1245 and 1423 cm$^2$V$^{-1}$s$^{-1}$, respectively. It should be noted that the quantum mobility (1249 cm$^2$V$^{-1}$s$^{-1}$) is comparable to the Hall mobility (1245 and 1423 cm$^2$V$^{-1}$s$^{-1}$).[38, 41] This is due to the fact that the Te vacancies become the scattering centers which significantly reduce the MR and the Hall mobility. Compared with our previous work (average mobility of ~730 cm$^2$V$^{-1}$s$^{-1}$ and carrier density of ~$10^{20}$ cm$^{-3}$),[30] the mobility here is almost doubled while the carrier density decreases in an order. This remarkable improvement of quality makes the more balanced *p-n* compensation and suppresses the relevant inelastic scatting of carriers, enabling the SdH oscillations emergent in the high-quality WTe$_2$ films.

In summary, we have fabricated the large-scale, high-quality, single-crystalline WTe$_2$ films on mica by the PLD and post-annealing method. The crystallization quality and the transport properties can be manipulated by tuning the annealing temperature and time. When the film is annealed at 700ºC for 72 hours, the distinct SdH oscillations are observed. Using the FFT and two-carriers model, physical quantities related to Fermi surface, such as the effective mass, nontrivial Berry phase, Dingle temperature, mobility and carrier density, are obtained. It is the doubled mobility and decreased carrier density that enable the observation of SdH quantum oscillations for the first time in the large-scale WTe$_2$ films. Our work raises the development of large-area topological film materials to a new stage, which provides a fertile land for the fabrication of the large-scale quantum devices based on Weyl materials.

This work was supported by the National Key R&D Program of China (Grant Nos. 2017YFA0206304 and 2016YFA0300803), the National Natural Science Foundation of China (Grant Nos 61822403, 11874203, 11774160, 61427812 and U1732159), the Fundamental Research Funds for the Central Universities (Grant Nos 021014380080 and 021014380113), the Natural Science Foundation of Jiangsu Province of China (Grant No BK20192006) and Collaborative Innovation Center of Solid-State Lighting and Energy-Saving Electronics.

**References**

[1] Schubnikow L and Haas W J d 1930 *Proc. Neth. R. Acad. Sci.* **33** 130
[2] D. Shoenberg 2009 *Magnetic Oscillations in Metals (Cambridge University Press, Cambridge, UK)*
[3] Chakravarty S and Kee H Y 2008 *Proc. Natl. Acad. Sci. USA* **105** 8835
[4] Tang F, Po H C, Vishwanath A et al 2019 *Nature* **566** 486
[5] Zhang T, Jiang Y, Song Z et al 2019 *Nature* **566** 475
[6] Vergniory M G, Elcoro L, Felser C et al 2019 *Nature* **566** 480
[7] Shekhar C, Nayak A K, Sun Y et al 2015 *Nat. Phys.* **11** 645
[8] Zhang H, Liu C-X, Qi X-L et al 2009 *Nat. Phys.* **5** 438
[9] Petrushevsky M, Lahoud E, Ron A et al 2012 *Phys. Rev. B* **86** 045131
[10] Xiang Z J, Zhao D, Jin Z et al 2015 *Phys. Rev. Lett.* **115** 226401
[11] Ali M N, Xiong J, Flynn S et al 2014 *Nature* **514** 205
[12] Wang C M, Lu H Z and Shen S Q 2016 *Phys. Rev. Lett.* **117** 077201
[13] Wang C M, Sun H P, Lu H Z et al 2017 *Phys. Rev. Lett.* **119** 136806
[14] Pletikosic I, Ali M N, Fedorov A V et al 2014 *Phys. Rev. Lett.* **113** 216601
[15] Kang D, Zhou Y, Yi W et al 2015 *Nat. Commun.* **6** 7804
[16] Pan X C, Chen X, Liu H et al 2015 *Nat. Commun.* **6** 7805
[17] Wang Y, Liu E, Liu H et al 2016 *Nat. Commun.* **7** 13142
[18] Fei Z Y, Palomaki T, Wu S F et al 2017 *Nat. Phys.* **13** 677
[19] Tang S J, Zhang C F, Wong D et al 2017 *Nat. Phys.* **13** 683
[20] Wu S F, Fatemi V, Gibson Q D et al 2018 *Science* **359** 76
[21] Fatemi V, Wu S F, Cao Y et al 2018 *Science* **362** 926
[22] Sajadi E, Palomaki T, Fei Z Y et al 2018 *Science* **362** 922
[23] Zhang E, Chen R, Huang C et al 2017 *Nano Lett.* **17** 878
[24] Zhou J, Lin J, Huang X et al 2018 *Nature* **556** 355
[25] Li J, Cheng S, Liu Z et al 2018 *J. Phys. Chem. C* **122** 7005
[26] Zhou J, Liu F, Lin J et al 2017 *Adv. Mater.* **29** 1603471
[27] Asaba T, Wang Y, Li G et al 2018 *Sci. Rep.* **8** 6520
[28] Jia Z-Y, Song Y-H, Li X-B et al 2017 *Phys. Rev. B* **96** 041108

[29] Yao J D, Zheng Z Q and Yang G W 2019 *Prog. Mater. Sci.* **106** 100573
[30] Gao M, Zhang M H, Niu W et al 2017 *Appl. Phys. Lett.* **111** 031906
[31] Vermeulen P A, Momand J and Kooi B J 2019 *Cryst. Eng. Comm.* **21** 3409
[32] Kong W D, F.Wu S, Richard P et al 2015 *Appl. Phys. Lett.* **106** 081906
[33] Cai P L, Hu J, He L P et al 2015 *Phys. Rev. Lett.* **115** 057202
[34] Lv Y Y, Zhang B B, Li X et al 2016 *Sci. Rep.* **6** 26903
[35] Wang X, Du Y, Dou S et al 2012 *Phys. Rev. Lett.* **108** 266806
[36] Zhang M, Wang X, Zhang S et al 2016 *IEEE Electron Dev. Lett.* **37** 1231
[37] Yang Y-K, Xiu F-X, Wang F-Q et al 2019 *Chin. Phys. B* **28** 107502
[38] Pan H, Zhang K, Wei Z et al 2016 *Appl. Phys. Lett.* **108** 183103
[39] Huang X-W, Liu X-X, Yu P et al 2019 *Chin. Phys. Lett.* **36** 077101
[40] Fatemi V, Gibson Q D, Watanabe K et al 2017 *Phys. Rev. B* **95** 041410
[41] Novak M, Sasaki S, Segawa K et al 2015 *Phys. Rev. B* **91** 041203

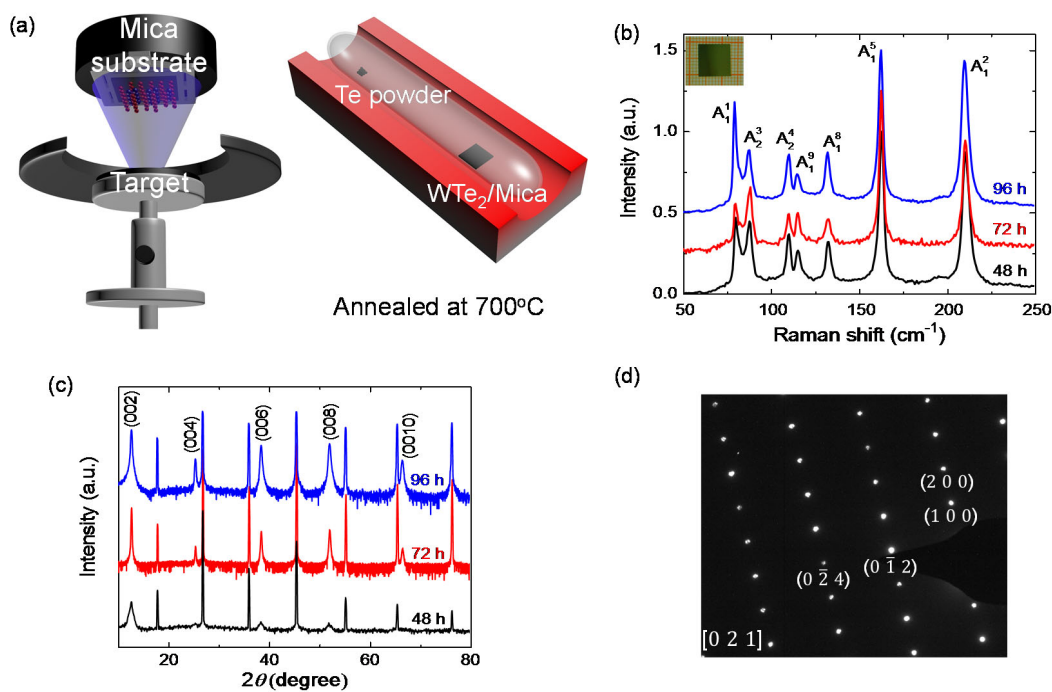

**Fig. 1.** Structural characterization of the WTe$_2$ films. (a) Schematic diagrams of the PLD and annealing apparatus. (b) Raman spectra of the WTe$_2$ films annealed for different time intervals. Inset shows the digital photograph of the centimeter-scale WTe$_2$ film. (c) XRD patterns of the WTe$_2$ films annealed for different time intervals. The WTe$_2$ film is characterized as (00*l*) family diffraction peaks while the other peaks belong to the mica substrate. (d) The SAED pattern taken along the [021] zone axis of the WTe$_2$ film annealed for 72 hours.

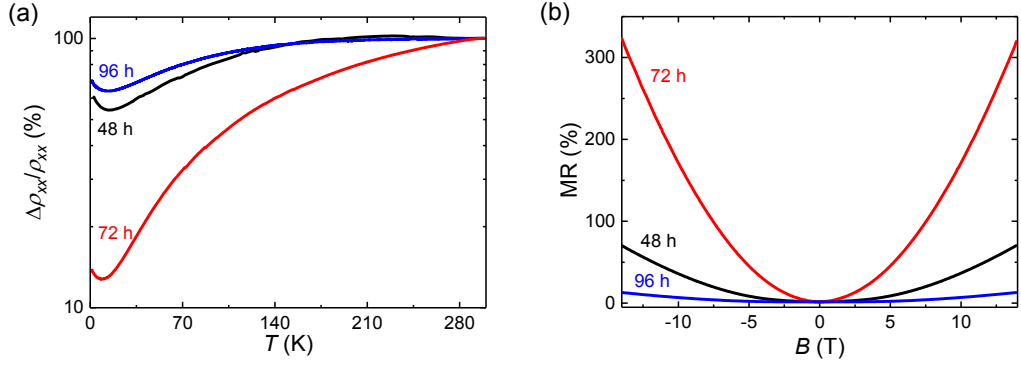

**Fig. 2.** (a) The temperature dependence of the longitudinal resistivity of the WTe$_2$ films annealed for different time intervals, as normalized by their values at 300 K. (b) The MR curves of the WTe$_2$ films annealed for different time intervals at 1.7 K.

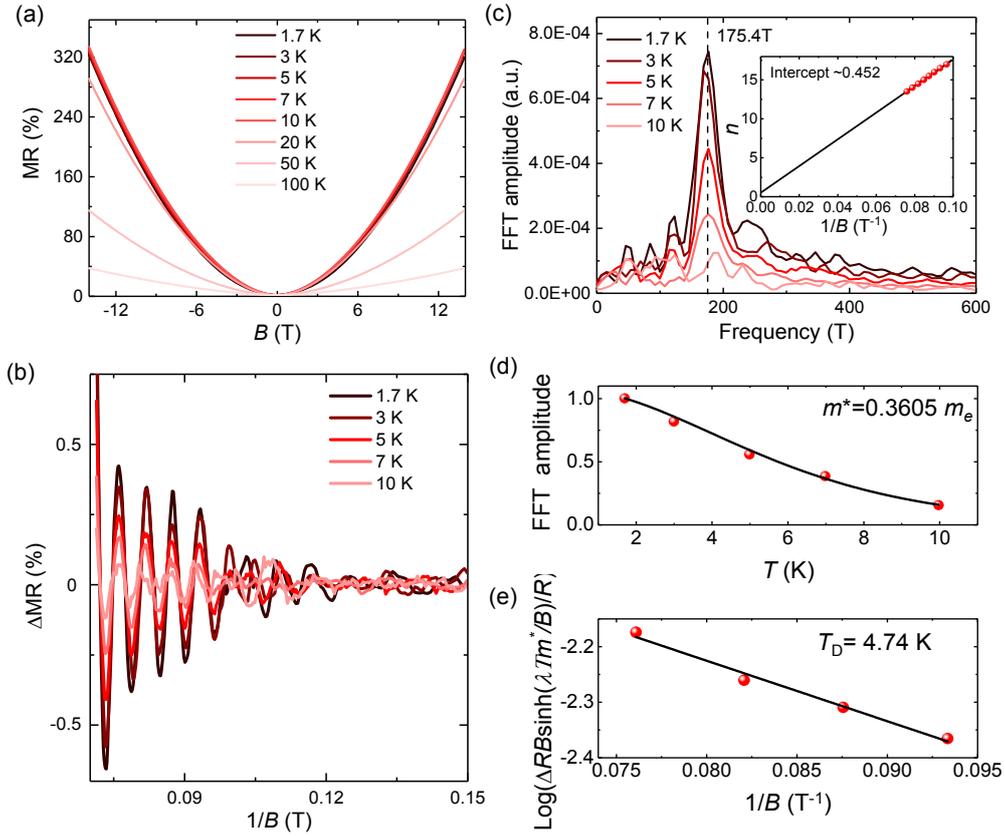

**Fig. 3.** (a) The MR curves of the WTe$_2$ film annealed for 72 hours at different temperatures. (b) Temperature dependence of SdH oscillations after a smooth subtraction of background. (c) The FFT of the SdH oscillations. Inset: the relationship between Landau level $n$ and the inverse field $1/B$. (d) The temperature dependence of FFT amplitude at 175.4 T, as normalized by the value at 1.7 K. (e) The Dingle plot of the $\mathrm{Log}(\Delta R \cdot B \cdot \sinh(\lambda T m^*/B)/R)$ versus the inverse field $1/B$ at 1.7 K.

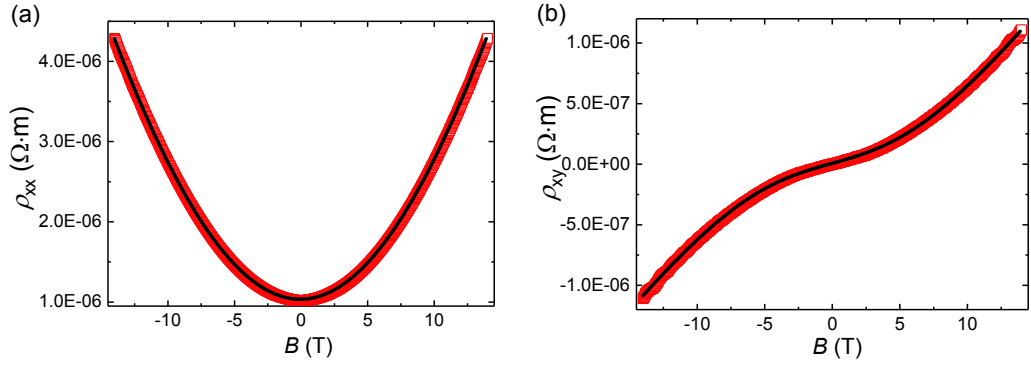

**Fig. 4.** Field-dependent longitudinal resistivity (a) and Hall resistivity (b) of the WTe$_2$ film annealed for 72 hours at 1.7 K. The black lines are the fitting curves by the two-carries model.

# Supplementary Materials:

# Observation of Shubnikov-de Haas Oscillations in Large-Scale Weyl Semimetal WTe$_2$ Films

Yequan Chen, Yongda Chen, Jiai Ning, Liming Chen, Wenzhuo Zhuang, Liang He, Rong Zhang, Yongbing Xu[*], Xuefeng Wang[*]

Jiangsu Provincial Key Laboratory of Advanced Photonic and Electronic Materials, School of Electronic Science and Engineering, and Collaborative Innovation Center of Advanced Microstructures, Nanjing University, Nanjing 210093, China

[*]Corresponding Authors. Email: xfwang@nju.edu.cn (X.W.); ybxu@nju.edu.cn (Y.X.)

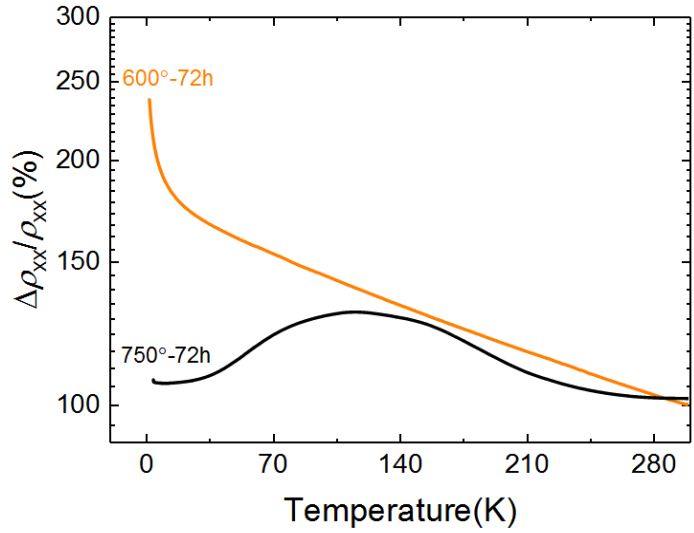

**Fig. S1.** Temperature dependent $\rho_{xx}$ of the WTe$_2$ films annealed at 600°C and 750°C for 72 hours, respectively. The $\rho_{xx}$ is normalized by the values measured at 300 K.

The temperature dependent $\rho_{xx}$ of the WTe$_2$ films annealed at 600°C and 750°C for 72 hours are shown in Fig. S1. The RRR values are much smaller than that annealed at 700°C [Fig. 2(a)]. The temperature dependent $\rho_{xx}$ even shows the insulating behavior, which further indicates that other annealing temperatures turn out the worse samples.

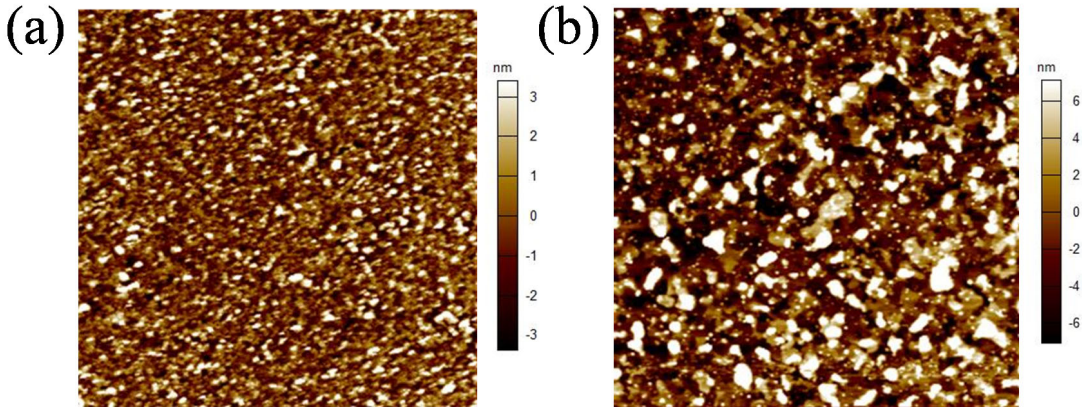

**Fig. S2.** 5×5 μm$^2$ surface morphology of the 100-nm film before (a) and after (b) annealing.

The roughness of the 100-nm film before annealing is ~1.5 nm; while the one after annealing is ~5 nm, as shown in Fig. S2. Here, the annealing process may slightly deteriorate the surface morphology of the WTe$_2$ film.